\documentclass[12pt]{article}
\usepackage[utf8]{inputenc}
\usepackage{authblk}
\usepackage{setspace}
\usepackage[margin=1.25in]{geometry}
\usepackage{graphicx}
\usepackage{subcaption}
\usepackage{amsmath}
\usepackage{lineno}
\usepackage{lipsum}
\usepackage[style=nejm, 
citestyle=numeric-comp,
sorting=none]{biblatex}
\begin{document}
\title{Recent highlights from the LHCb experiment}
\author[1*]{Abhijit Mathad (on behalf of the LHCb collaboration)}
\affil[1]{Institute Physik, Universit\"at Z\"urich, Switzerland}

\onehalfspacing
\maketitle

\date{}

\begin{abstract}
  The Large Hadron Collider beauty (LHCb) detector is a single-arm forward spectrometer at the LHC, designed for the study of heavy flavour physics. 
  In this proceedings, 
  an overview of the detector performance and a few recent results in the field of beauty and charm physics are presented. The LHCb experiment has also undergone a major upgrade in preparation for Run 3 and Run 4 of LHC, the proceedings will briefly highlight the current activities undertaken during this period. 
  This proceedings is based on the plenary talk presented at the XXV DAE-BRNS High Energy Physics Symposium 
  held from 12-16 December 2022 at IISER Mohali, India.
\end{abstract}

\section{Introduction}

The Large Hadron Collider beauty (LHCb) detector, located near Ferney-Voltaire, France, operates at the LHC interaction point 8 (IP8) at CERN. Originally proposed in 1998~\cite{CERN-LHCC-98-004,Belyaev:2021cyr}, LHCb has evolved into a forward general purpose detector, exploring beauty and charm hadron decays with exceptional precision. It plays a crucial role in measuring flavor observables to probe new physics beyond the Standard Model at multi-TeV energy scales~\cite{UTfit:2007eik,Artuso:2022ouk}. 

The proceedings are organised into sections that describe the LHCb detector and its performance (section~\ref{sec:detector}), recent highlights from the experiment (section~\ref{sec:physics}), and a brief overview of the LHCb upgrade (section~\ref{sec:conclusion}).

\section{Detector and performance}
\label{sec:detector}

The LHCb detector is a single-arm forward spectrometer focused on studying beauty and charm hadron decays in $pp$ collisions at the LHC~\cite{LHCb:2008vvz}. It benefits from the abundant production of $b\bar{b}$ and $c\bar{c}$ pairs, with cross-sections $\sigma_{b\bar{b}}\approx 500,\mu$b and $\sigma_{c\bar{c}}\approx 20 \times \sigma_{b\bar{b}}$ at $\sqrt{s}=13$ TeV. LHCb observes all species of beauty and charm hadrons produced and boosted forward, distinguishing it from B factories. 

The detector includes a high-precision track and vertex reconstruction system, a particle identification system, and an efficient trigger system. The tracking and vertex information come from the VErtex LOcator (VELO), silicon-strip and straw tube trackign detectors (TT and T1-T3), and multi-wire proportional chambers in the muon system. The VELO reconstructs primary and secondary vertices, while the tracking system achieves excellent momentum resolution ($\sigma_{p_T}/p_T\approx 0.5\%$ for $p_T=10$~GeV/$c$) and decay time resolution ($\sigma_t\approx 45$~fs for $\tau_{B^0_s}\approx 1.5$~ps)~\cite{LHCb:2014set}.
Particle identification relies on RICH detectors (RICH1 and RICH2) and the calorimeter system (SPD, PS, ECAL, and HCAL) in conjunction with the muon system. The PID system demonstrates efficient separation between pions, kaons, and protons, with a kaon identification efficiency $> 95\%$ and a pion mis-identification rate $< 5\%$.
The LHCb trigger system consists of a hardware trigger (Level-0) and a software trigger (HLT), reducing the event rate from 40MHz to 1MHz and further to 12.5~kHz, respectively~\cite{Aaij:2019uij,Aaij:2016rxn}.

Data collection at LHCb occurred at proton-proton center-of-mass energies of 7, 8, and 13 TeV during 2010-2011, 2012, and 2015-2018, with integrated luminosities of $1.15,\rm{fb}^{-1}$, $2.08,\rm{fb}^{-1}$, and $5.9,\rm{fb}^{-1}$, respectively.

\section{Highlights of physics results}
\label{sec:physics}

The LHCb collaboration has published over 600 papers spanning 
a broad range of topics such as studies of  
Charge-Parity (CP) violation
in beauty and charm decays, semileptonic decays, spectroscopy, electroweak physics, and heavy-ion physics.
In this proceedings, we will discuss results concerning 
charm physics, semileptonic decays, and CP violation in beauty decays.

\subsection{CP violation and charm physics}
\label{sec:charm}

The Cabibbo-Kobayashi-Maskawa (CKM) matrix within the Standard Model governs quark flavor mixing and is represented by a unitary $3\times 3$ matrix, satisfying one of the unitarity conditions $V_{ud}V_{ub}^* + V_{cd}V_{cb}^* + V_{td}V_{tb}^* = 0$.
This unitarity condition can be represented as a triangle in the complex plane, known as the Unitarity Triangle (UT)~\cite{UTfit:2007eik}.
Remarkable progress has been made since 1995 through measurements of angles and sides of the UT. A pivotal measurement is the angle $\gamma$, denoting the phase of the CKM matrix element $V_{ub}$. Obtained from tree-level $B$ meson decays, $\gamma$ can be compared with its indirect estimate from the UT fit, utilising observables sensitive to loop-level processes. This rigorous examination serves as a stringent test of the SM and a probe for new physics (NP) effects.

\begin{figure}[!htp]
  \centering
  \includegraphics[width=0.7\linewidth]{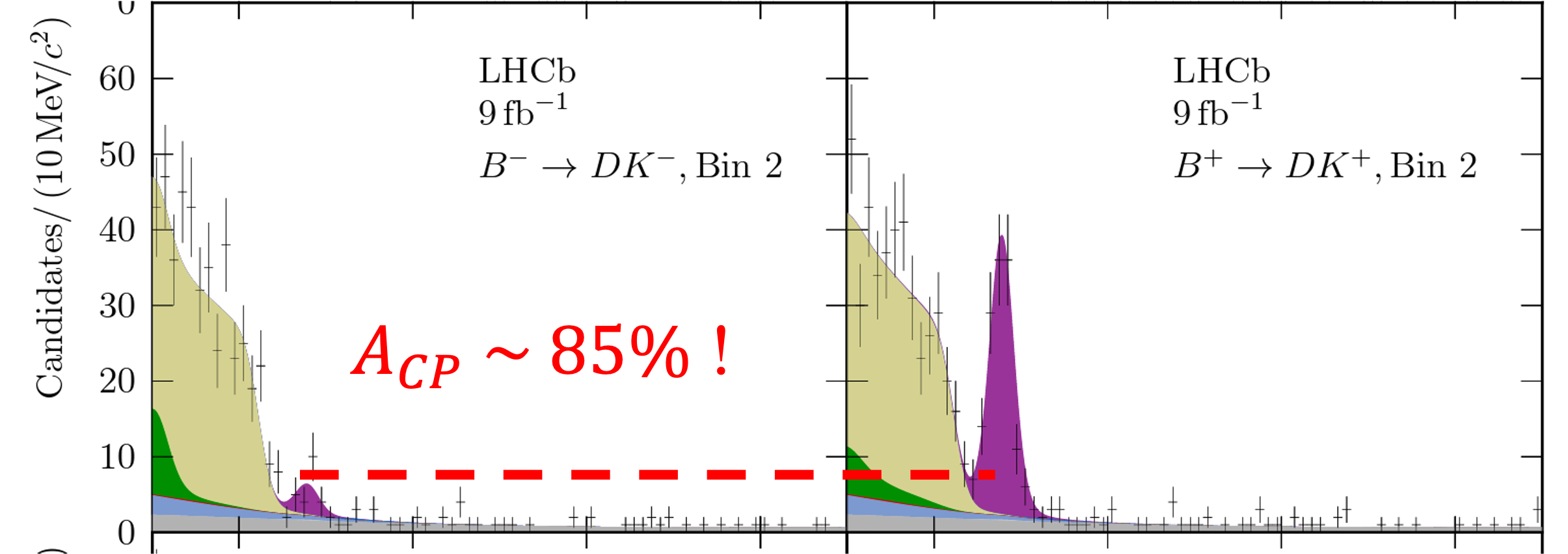}
  \includegraphics[width=0.45\linewidth]{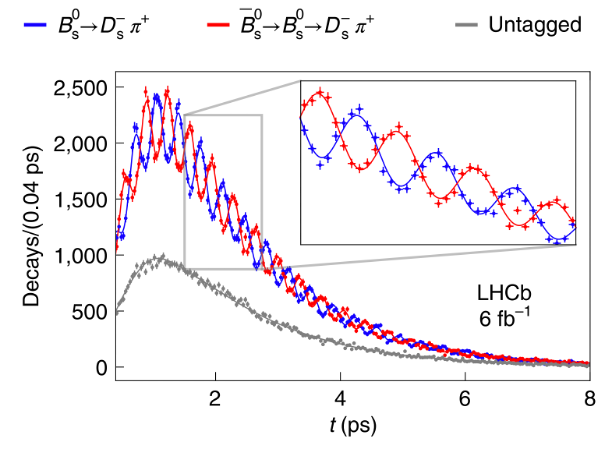}
  \includegraphics[width=0.4\linewidth]{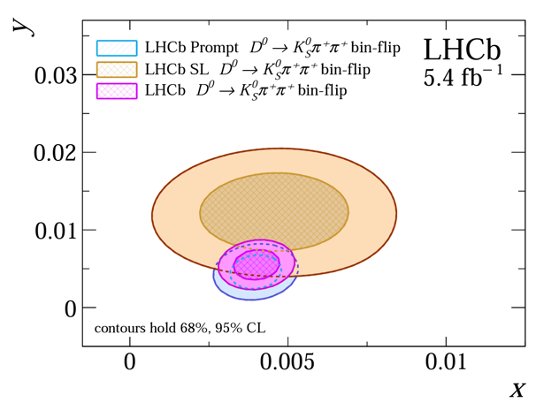}
  \caption{
  (Top) Asymmetry between $B^\pm\to D K^\pm$ and its charge conjugate decay used to extract $\gamma$.
  (Bottom-left) The time distribution of $B_s$ mesons decaying to $D_s^- \pi^+$. (Bottom-right) The measurement of charm mixing parameters $x$ and $y$.}
  \label{fig:gamma1}
\end{figure}

In a recent paper by LHCb~\cite{LHCb:2022nng}, the angle $\gamma$ was precisely measured via $B^\pm\to D K^\pm$ decays, with $D$ being either $D^0$ or $\bar{D}^0$ mesons reconstructed from $D\to K^\mp\pi^\pm\pi^\pm\pi^\mp$ decays. The model-independent extraction of $\gamma$ was performed in four phase space bins of $D$ decay using Run 1 and Run 2 data. The analysis explored the ratio of opposite-sign to like-sign decay channels to obtain $\gamma$. Charm decay parameters were obtained from external input from CLEO-c and BESIII.
Figure~\ref{fig:gamma1} shows the asymmetry between $B^\pm\to D K^\pm$ and its charge conjugate decay, which is used in the extraction of angle $\gamma$.
The precise measurement resulted in 
$\gamma = (54.8^{+6.0}_{-5.8}$$^{+0.6}_{-0.6}$$^{+6.7}_{-4.3})^\circ$,
with uncertainties attributed to statistical, systematic, and external input sources. This measurement represents the most precise determination of $\gamma$ to date using a single $D$ decay mode.

Another essential observable utilised to test the unitarity of the CKM matrix is the oscillation frequency ($\Delta m_s$) of $B_s$ mesons. A recent LHCb measurement investigates $B_s$ meson oscillation to $\bar{B}_s$ mesons via $B_s\to D_s^- \pi^+$ decays~\cite{LHCb:2022nng}. This measurement utilises Run 1 and Run 2 data, reconstructing $D_s^-$ from $D_s^-\to K^+K^-\pi^-$ and $D_s^-\to \pi^+\pi^-\pi^-$ decays. The oscillation frequency ($\Delta m_s$) is extracted from the time distribution of $B_s$ mesons, as shown in Figure~\ref{fig:gamma1} on the bottom-left. The measured value of $\Delta m_s$ is $17.7683\pm0.0051\pm0.0032,\rm{ps}^{-1}$, where the first uncertainty is statistical, and the second is systematic. This measurement represents the most precise determination of $\Delta m_s$ to date.

Particle-antiparticle oscillation can occur in the charm sector with mixing parameters: $x=\Delta m/\Gamma$ and $y= \Delta\Gamma/2\Gamma$. LHCb made the first observation of $D^0$ oscillations in 2012~\cite{LHCb:2012zll}, and the first observation of $x\neq 0$ was made in 2021~\cite{LHCb:2021ykz}. A novel study, utilising full Run 2 data, analyses the time evolution of $D^0\to K_S \pi^+ \pi^-$ phase space, requiring $D^0$ to originate from $B^0\to D^0 \mu^-\bar{\nu}_\mu X$ decays~\cite{DiCanto:2018tsd,LHCb:2022rdf}. This analysis extracts mixing and CPV parameters by measuring yield ratios between different $D^0$ phase space bins as a function of the $D^0$ decay time. The result of the measurement 
is shown in Figure~\ref{fig:gamma1} on the bottom-right. 
Uncertainty on external inputs from CLEO and BESIII are the limiting systematic for this measurement. 

A new LHCb combination of observables yields $\gamma=(63.8_{-3.7}^{+3.5})^\circ$ and charm mixing parameters from 173 measurements. This value is consistent with the indirect estimate $\gamma=(65.5_{-2.7}^{+1.1})^\circ$ and extractions from UTfit and CKMfitter. However, tensions between different B species for $\gamma$ persist at around 2 sigma~\cite{LHCb-CONF-2022-003}. 

\subsection{Semileptonic decays}
\label{sec:semileptonic}

The semileptonic decays of $b$-hadrons can proceed via two types of transitions: charged current ($b\to c\ell\nu$) and neutral current ($b\to s\ell^+\ell^-$). The charged current decays (CC) have relatively large branching fractions of a few percent and yield significant signals, but they pose experimental challenges due to the presence of undetectable neutrinos. As a result, their measurements are dominated by systematic uncertainties. On the other hand, the neutral current decays (NC) occur only through loop processes and have branching fractions of approximately $O(10^{-7})$. These measurements are typically limited by statistical uncertainties, but they offer opportunities to explore the effects of new particles that might enhance amplitudes suppressed in the Standard Model.

\begin{figure}[!htp]
  \centering
  \includegraphics[width=0.47\linewidth]{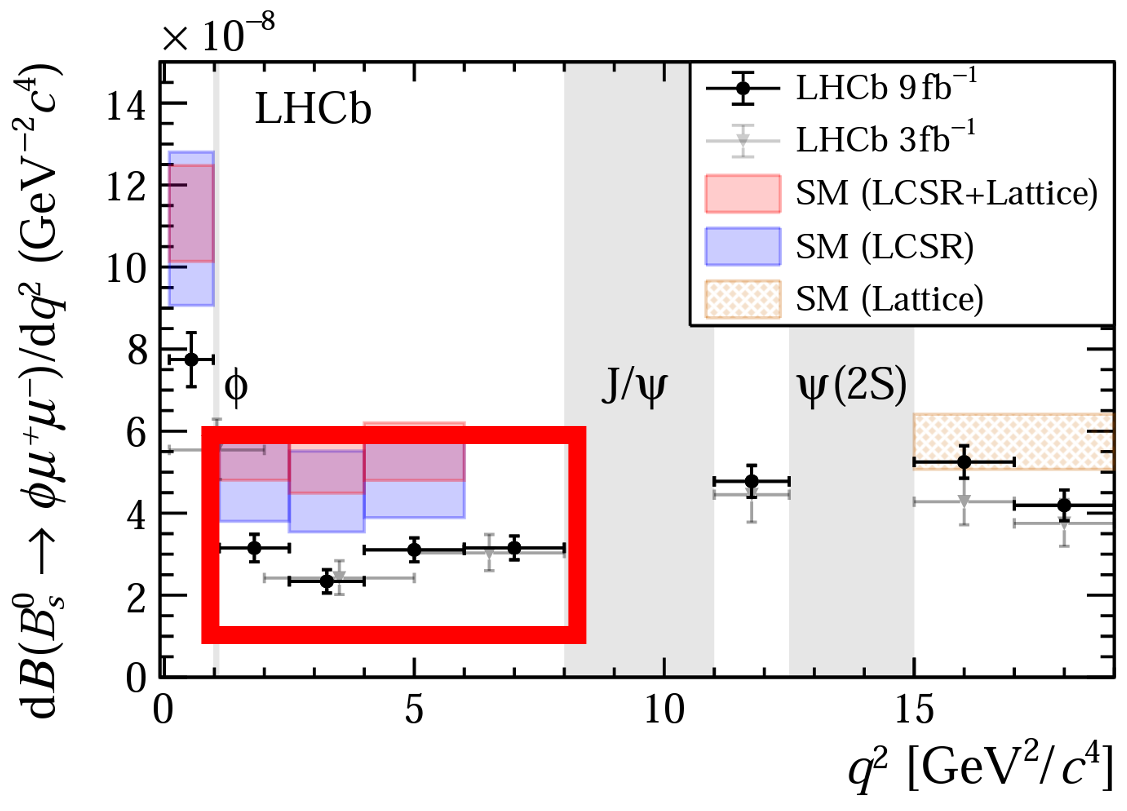}
  \includegraphics[width=0.47\linewidth]{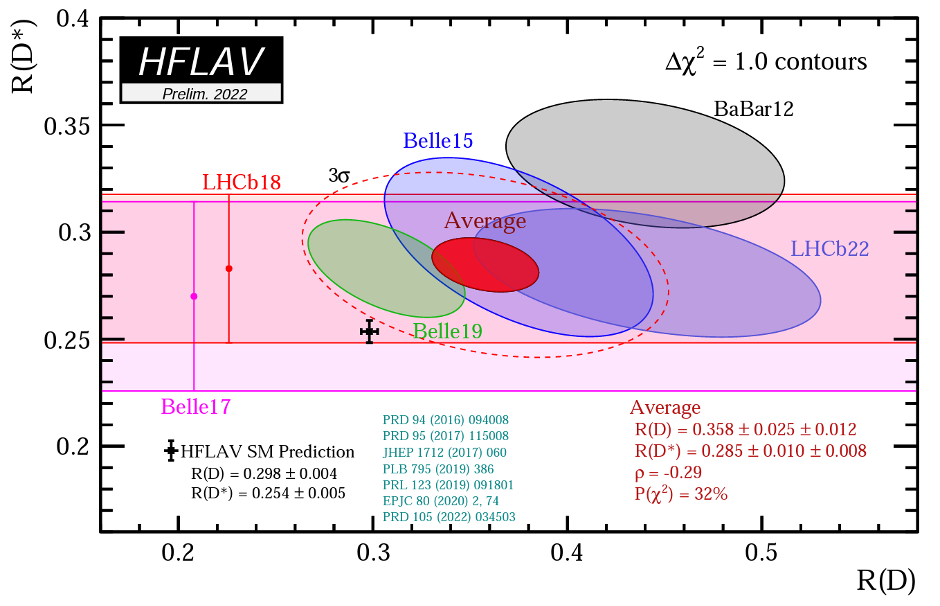}
  \caption{
  (Left) Different measurement of $B_s^0\to\phi\mu^+\mu^-$ branching fraction in neutral current decays.
  (Right) Measurement of lepton flavour universality ratio 
  $R(D)$-$R(D^*)$ in charged current decays.}
  \label{fig:semileptonic}
\end{figure}

The LHCb experiment has undertaken various measurements related to both NC and CC semileptonic decays. In the context of NC decays, LHCb has investigated pure leptonic decays of $B$ mesons, which are theoretically and experimentally cleaner processes. The most recent measurement reveals the branching fraction for $B_s^0\to\mu^+\mu^-$ to be $(3.9_{-0.43-0.11}^{+0.46+0.15})\times10^{-9}$ and constrains $BF(B^0 \to \mu^+ \mu^-)$ to be less than $2.6\times10^{-10}$~\cite{LHCb:2021vsc}. Additionally, differential measurements of NC decays, particularly $\frac{d\Gamma}{dq^2}$ (where $q^2=m^2(l^+l^-)$, excluding $c\bar{c}$ resonances), have been conducted~\cite{LHCb:2015tgy,LHCb:2014cxe,LHCb:2016ykl}. Notably, the recent measurement for $B_s^0\to\phi\mu^+\mu^-$ shows a $3.6\sigma$ tension with respect to the Standard Model, indicating potential hints of new physics, as depicted in Figure~\ref{fig:semileptonic} on the left~\cite{LHCb:2021zwz}. Moreover, angular observables, including $P_5\prime$ and $F_L$, have been measured, showing consistent trends across different modes, hinting at the presence of a new physics vector coupling with $Re(\Delta C_9)\approx -1$~\cite{Albrecht:2021tul,LHCb:2021xxq,LHCb:2020gog,LHCb:2020lmf}.

LHCb has also conducted tests of lepton flavor universality (LFU) within the Standard Model for both NC and CC decays. LFU implies that the ratio of branching fractions for decays involving different species of leptons should only depend on the masses of the leptons. For NC decays, the ratios $R_K$ and $R_{K^*}$, which involve muons and electrons, have been measured and found to be consistent with the Standard Model~\cite{LHCb:2021trn}
\footnote{
Since the presentation of this talk, a new measurement of $R_K$ and $R_{K^*}$ has been published~\cite{LHCb:2022zom}, superseding the previous reference~\cite{LHCb:2021trn}.
}. Similarly, for CC decays, LHCb has carried out its first simultaneous measurement of the ratios $R(D)$ and $R(D^*)$, which involves $\tau$ and $\mu$ leptons. The results from data collected during the 2016 run period yield $R(D)=0.441\pm0.060\pm0.066$ and $R(D^*)=0.281\pm0.018\pm0.023$, with a correlation coefficient of $-0.43$~\cite{LHCb:2023zxo}. Figure~\ref{fig:semileptonic} on the right compares these results with other measurements and Standard Model predictions. Notably, the world average of $R(D)$ and $R(D^*)$ exhibits a $3.3\sigma$ tension with respect to the Standard Model
\footnote{
Since this talk, two new measurements of $R(D)$ and $R(D^*)$ have been presented. One by LHCb~\cite{LHCb:2023cjr} and the other by Belle-II~\cite{Belletwo}. However, the significance of the tension with respect to SM remains unchanged.
}.

\section{Future prospects and conclusion}
\label{sec:conclusion}

The LHCb experiment is currently undergoing a significant upgrade to its detector, in preparation for Run 3 of the LHC. This upgrade is tailored to handle a luminosity that is five times higher than that of Run 2. The detector enhancements include a cutting-edge pixel-based vertex locator, an improved tracking system with higher granularity, upgraded RICH detectors equipped with new photon detectors, and a state-of-the-art readout system. Moreover, the LHCb trigger system is transitioning to a fully software-based approach, enabling more flexibility and efficiency in triggering processes. With these improvements, LHCb aims to amass an impressive data sample of 50 fb$^{-1}$ by the conclusion of LHC Run 4, and an astounding 300 fb$^{-1}$ during the operation of the High Luminosity LHC (HL-LHC). In the HL-LHC era, the instantaneous luminosity will surge by a factor of 50, and ongoing research and development are actively exploring the implementation of new detector components with timing information. The expanded data set will empower LHCb to achieve more precise measurements of rare decays and CP violation phenomena, while also providing the capability to thoroughly search for new physics across a wide array of channels. The LHCb upgrade heralds an exciting era of particle physics exploration, unlocking deeper insights into the fundamental properties of the universe.

\printbibliography

@book{CERN-LHCC-98-004,
      title         = "{LHCb : Technical Proposal}",
      publisher     = "CERN",
      address       = "Geneva",
      year          = "1998",
      url           = "https://cds.cern.ch/record/622031",
}

@article{Belyaev:2021cyr,
    author = "Belyaev, I. and Carboni, G. and Harnew, N. and Matteuzzi, C. and Teubert, F.",
    title = "{The history of LHCb}",
    eprint = "2101.05331",
    archivePrefix = "arXiv",
    primaryClass = "physics.hist-ph",
    doi = "10.1140/epjh/s13129-021-00002-z",
    journal = "Eur. Phys. J. H",
    volume = "46",
    number = "1",
    pages = "3",
    year = "2021"
}

@article{UTfit:2007eik,
    author = "Bona, M. and others",
    collaboration = "UTfit",
    title = "{Model-independent constraints on $\Delta F=2$ operators and the scale of new physics}",
    eprint = "0707.0636",
    archivePrefix = "arXiv",
    primaryClass = "hep-ph",
    doi = "10.1088/1126-6708/2008/03/049",
    journal = "JHEP",
    volume = "03",
    pages = "049",
    year = "2008"
}

@article{Artuso:2022ouk,
    author = "Artuso, Marina and others",
    title = "{Report of the Frontier For Rare Processes and Precision Measurements}",
    eprint = "2210.04765",
    archivePrefix = "arXiv",
    primaryClass = "hep-ex",
    reportNumber = "FERMILAB-FN-1206-PPD",
    month = "10",
    year = "2022"
}

@article{LHCb:2008vvz,
    author = "Alves, Jr., A. Augusto and others",
    collaboration = "LHCb",
    title = "{The LHCb Detector at the LHC}",
    reportNumber = "LHCb-DP-2008-001",
    doi = "10.1088/1748-0221/3/08/S08005",
    journal = "JINST",
    volume = "3",
    pages = "S08005",
    year = "2008"
}

@article{LHCb:2014set,
    author = "Aaij, Roel and others",
    collaboration = "LHCb",
    title = "{LHCb Detector Performance}",
    eprint = "1412.6352",
    archivePrefix = "arXiv",
    primaryClass = "hep-ex",
    reportNumber = "LHCB-DP-2014-002, CERN-PH-EP-2014-290",
    doi = "10.1142/S0217751X15300227",
    journal = "Int. J. Mod. Phys. A",
    volume = "30",
    number = "07",
    pages = "1530022",
    year = "2015"
}

@article{Aaij:2019uij,
    author = "Aaij, R. and others",
    title = "{A comprehensive real-time analysis model at the LHCb experiment}",
    eprint = "1903.01360",
    archivePrefix = "arXiv",
    primaryClass = "hep-ex",
    reportNumber = "CERN-LHCb-DP-2019-002",
    doi = "10.1088/1748-0221/14/04/P04006",
    journal = "JINST",
    volume = "14",
    number = "04",
    pages = "P04006",
    year = "2019"
}

@article{Aaij:2016rxn,
    author = "Aaij, R. and others",
    title = "{Tesla : an application for real-time data analysis in High Energy Physics}",
    eprint = "1604.05596",
    archivePrefix = "arXiv",
    primaryClass = "physics.ins-det",
    reportNumber = "CERN-LHCB-DP-2016-001",
    doi = "10.1016/j.cpc.2016.07.022",
    journal = "Comput. Phys. Commun.",
    volume = "208",
    pages = "35--42",
    year = "2016"
}

@article{LHCb:2022nng,
    author = "Aaij, R. and others",
    collaboration = "LHCb",
    title = "{Measurement of the CKM angle $\gamma$ with $ B^\pm \to D[K^\mp \pi^\pm \pi^\pm \pi^\mp] h^\pm$ decays using a binned phase-space approach}",
    eprint = "2209.03692",
    archivePrefix = "arXiv",
    primaryClass = "hep-ex",
    reportNumber = "LHCb-PAPER-2022-017, CERN-EP-2022-150",
    doi = "10.1007/JHEP07(2023)138",
    journal = "JHEP",
    volume = "07",
    pages = "138",
    year = "2023"
}

@article{DiCanto:2018tsd,
    author = "Di Canto, A. and Garra Tic\'o, J. and Gershon, T. and Jurik, N. and Martinelli, M. and Pila\v{r}, T. and Stahl, S. and Tonelli, D.",
    title = "{Novel method for measuring charm-mixing parameters using multibody decays}",
    eprint = "1811.01032",
    archivePrefix = "arXiv",
    primaryClass = "hep-ex",
    doi = "10.1103/PhysRevD.99.012007",
    journal = "Phys. Rev. D",
    volume = "99",
    number = "1",
    pages = "012007",
    year = "2019"
}

@article{LHCb:2022rdf,
    collaboration = "LHCb",
    title = "{Model-independent measurement of charm mixing parameters in $\bar{B} \rightarrow D^0 ( \rightarrow K_S^0 \pi^+ \pi^-) \mu^- \bar{\nu}_\mu X$ decays}",
    eprint = "2208.06512",
    archivePrefix = "arXiv",
    primaryClass = "hep-ex",
    reportNumber = "LHCb-PAPER-2022-020, CERN-EP-2022-160",
    month = "8",
    year = "2022"
}

@article{LHCb:2012zll,
    author = "Aaij, R and others",
    collaboration = "LHCb",
    title = "{Observation of $D^0 - \overline{D}^0$ oscillations}",
    eprint = "1211.1230",
    archivePrefix = "arXiv",
    primaryClass = "hep-ex",
    reportNumber = "CERN-PH-EP-2012-333, LHCB-PAPER-2012-038",
    doi = "10.1103/PhysRevLett.110.101802",
    journal = "Phys. Rev. Lett.",
    volume = "110",
    number = "10",
    pages = "101802",
    year = "2013"
}

@article{LHCb:2021ykz,
    author = "Aaij, Roel and others",
    collaboration = "LHCb",
    title = "{Observation of the Mass Difference Between Neutral Charm-Meson Eigenstates}",
    eprint = "2106.03744",
    archivePrefix = "arXiv",
    primaryClass = "hep-ex",
    reportNumber = "LHCb-PAPER-2021-009, CERN-EP-2021-099",
    doi = "10.1103/PhysRevLett.127.111801",
    journal = "Phys. Rev. Lett.",
    volume = "127",
    number = "11",
    pages = "111801",
    year = "2021"
}

@techreport{LHCb-CONF-2022-003,
      collaboration = "LHCb",
      title         = "{Simultaneous determination of the CKM angle $\gamma$ and
                       parameters related to mixing and CP violation in the charm
                       sector}",
      institution   = "CERN",
      reportNumber  = "LHCb-CONF-2022-003, CERN-LHCb-CONF-2022-003",
      address       = "Geneva",
      year          = "2022",
      url           = "https://cds.cern.ch/record/2838029",
}

@article{LHCb:2021vsc,
    author = "Aaij, R. and others",
    collaboration = "LHCb",
    title = "{Analysis of Neutral B-Meson Decays into Two Muons}",
    eprint = "2108.09284",
    archivePrefix = "arXiv",
    primaryClass = "hep-ex",
    reportNumber = "CERN-CERN-EP-2021-132, LHCb-PAPER-2021-007",
    doi = "10.1103/PhysRevLett.128.041801",
    journal = "Phys. Rev. Lett.",
    volume = "128",
    number = "4",
    pages = "041801",
    year = "2022"
}

@article{LHCb:2021zwz,
    author = "Aaij, Roel and others",
    collaboration = "LHCb",
    title = "{Branching Fraction Measurements of the Rare $B^0_s\rightarrow\phi\mu^+\mu^-$ and $B^0_s\rightarrow f_2(1525)\mu^+\mu^-$- Decays}",
    eprint = "2105.14007",
    archivePrefix = "arXiv",
    primaryClass = "hep-ex",
    reportNumber = "LHCb-PAPER-2021-014, CERN-EP-2021-092",
    doi = "10.1103/PhysRevLett.127.151801",
    journal = "Phys. Rev. Lett.",
    volume = "127",
    number = "15",
    pages = "151801",
    year = "2021"
}

@article{LHCb:2015tgy,
    author = "Aaij, Roel and others",
    collaboration = "LHCb",
    title = "{Differential branching fraction and angular analysis of $\Lambda^{0}_{b} \rightarrow \Lambda \mu^+\mu^-$ decays}",
    eprint = "1503.07138",
    archivePrefix = "arXiv",
    primaryClass = "hep-ex",
    reportNumber = "LHCB-PAPER-2015-009, CERN-PH-EP-2015-078",
    doi = "10.1007/JHEP06(2015)115",
    journal = "JHEP",
    volume = "06",
    pages = "115",
    year = "2015",
    note = "[Erratum: JHEP 09, 145 (2018)]"
}

@article{LHCb:2014cxe,
    author = "Aaij, R. and others",
    collaboration = "LHCb",
    title = "{Differential branching fractions and isospin asymmetries of $B \to K^{(*)} \mu^+ \mu^-$ decays}",
    eprint = "1403.8044",
    archivePrefix = "arXiv",
    primaryClass = "hep-ex",
    reportNumber = "LHCB-PAPER-2014-006, CERN-PH-EP-2014-055",
    doi = "10.1007/JHEP06(2014)133",
    journal = "JHEP",
    volume = "06",
    pages = "133",
    year = "2014"
}

@article{LHCb:2016ykl,
    author = "Aaij, Roel and others",
    collaboration = "LHCb",
    title = "{Measurements of the S-wave fraction in $B^{0}\rightarrow K^{+}\pi^{-}\mu^{+}\mu^{-}$ decays and the $B^{0}\rightarrow K^{\ast}(892)^{0}\mu^{+}\mu^{-}$ differential branching fraction}",
    eprint = "1606.04731",
    archivePrefix = "arXiv",
    primaryClass = "hep-ex",
    reportNumber = "CERN-EP-2016-141, LHCB-PAPER-2016-012",
    doi = "10.1007/JHEP11(2016)047",
    journal = "JHEP",
    volume = "11",
    pages = "047",
    year = "2016",
    note = "[Erratum: JHEP 04, 142 (2017)]"
}

@article{Albrecht:2021tul,
    author = "Albrecht, Johannes and van Dyk, Danny and Langenbruch, Christoph",
    title = "{Flavour anomalies in heavy quark decays}",
    eprint = "2107.04822",
    archivePrefix = "arXiv",
    primaryClass = "hep-ex",
    doi = "10.1016/j.ppnp.2021.103885",
    journal = "Prog. Part. Nucl. Phys.",
    volume = "120",
    pages = "103885",
    year = "2021"
}

@article{LHCb:2021xxq,
    author = "Aaij, Roel and others",
    collaboration = "LHCb",
    title = "{Angular analysis of the rare decay $B^0_s \rightarrow \phi \mu^+ \mu^-$}",
    eprint = "2107.13428",
    archivePrefix = "arXiv",
    primaryClass = "hep-ex",
    reportNumber = "LHCb-PAPER-2021-022, CERN-EP-2021-138",
    doi = "10.1007/JHEP11(2021)043",
    journal = "JHEP",
    volume = "11",
    pages = "043",
    year = "2021"
}

@article{LHCb:2020gog,
    author = "Aaij, Roel and others",
    collaboration = "LHCb",
    title = "{Angular Analysis of the  $B^{+}\rightarrow K^{*+}\mu^{+}\mu^{-}$ Decay}",
    eprint = "2012.13241",
    archivePrefix = "arXiv",
    primaryClass = "hep-ex",
    reportNumber = "LHCb-PAPER-2020-041, CERN-EP-2020-239",
    doi = "10.1103/PhysRevLett.126.161802",
    journal = "Phys. Rev. Lett.",
    volume = "126",
    number = "16",
    pages = "161802",
    year = "2021"
}

@article{LHCb:2020lmf,
    author = "Aaij, Roel and others",
    collaboration = "LHCb",
    title = "{Measurement of $CP$-Averaged Observables in the $B^{0}\rightarrow K^{*0}\mu^{+}\mu^{-}$ Decay}",
    eprint = "2003.04831",
    archivePrefix = "arXiv",
    primaryClass = "hep-ex",
    reportNumber = "LHCb-PAPER-2020-002, CERN-EP-2020-027",
    doi = "10.1103/PhysRevLett.125.011802",
    journal = "Phys. Rev. Lett.",
    volume = "125",
    number = "1",
    pages = "011802",
    year = "2020"
}

@article{LHCb:2021trn,
    author = "Aaij, Roel and others",
    collaboration = "LHCb",
    title = "{Test of lepton universality in beauty-quark decays}",
    eprint = "2103.11769",
    archivePrefix = "arXiv",
    primaryClass = "hep-ex",
    reportNumber = "LHCb-PAPER-2021-004, CERN-EP-2021-042",
    doi = "10.1038/s41567-021-01478-8",
    journal = "Nature Phys.",
    volume = "18",
    number = "3",
    pages = "277--282",
    year = "2022"
}

@article{LHCb:2022zom,
    collaboration = "LHCb",
    title = "{Measurement of lepton universality parameters in $B^+\to K^+\ell^+\ell^-$ and $B^0\to K^{*0}\ell^+\ell^-$ decays}",
    eprint = "2212.09153",
    archivePrefix = "arXiv",
    primaryClass = "hep-ex",
    month = "12",
    year = "2022"
}

@article{LHCb:2023cjr,
    author = "Aaij, Roel and others",
    collaboration = "LHCb",
    title = "{Test of lepton flavour universality using $B^0 \to D^{*-}\tau^+\nu_{\tau}$ decays with hadronic $\tau$ channels}",
    eprint = "2305.01463",
    archivePrefix = "arXiv",
    primaryClass = "hep-ex",
    reportNumber = "LHCb-PAPER-2022-052, CERN-EP-2023-062",
    month = "5",
    year = "2023"
}

@article{LHCb:2023zxo,
    collaboration = "LHCb",
    title = "{Measurement of the ratios of branching fractions $\mathcal{R}(D^{*})$ and $\mathcal{R}(D^{0})$}",
    eprint = "2302.02886",
    archivePrefix = "arXiv",
    primaryClass = "hep-ex",
    reportNumber = "LHCb-PAPER-2022-039, CERN-EP-2022-284",
    month = "2",
    year = "2023"
}

@misc{Belletwo,
    author = "Kazuki Kojima for Belle-II Collaboration, Talk at Lepton Photon 2023",
    title = "{Test of lepton flavour universality using $B^0 \to D^{*-}\tau^+\nu_{\tau}$ decays with hadronic $\tau$ channels}",
    year = "2023",
    URL="https://indico.cern.ch/event/1114856/contributions/5423684/attachments/2685890/4660084/2023-07-04_LP2023_KojimaFinalVer_main.pdf",
}
\end{document}